\documentclass[cup6b]{cupbook}

\usepackage{epsf}
\newcommand{\dblspace}{}

\def\note #1]{{\bf #1]}}

\def\fig{.}

\def\note #1]{{\bf #1]}}
\def\dd{{\rm d}}
\def\be{{\bf e}}
\def\bu{{\bf u}}
\def\bx{{\bf x}}
\def\CK{{\cal K}}
\def\rt{r_{\rm t}}
\def\PP{{\cal P}}
\def\Eq #1{Eq.\ (\ref{#1})}
\author{J{\O}RGEN CHRISTENSEN-DALSGAARD\\
Teoretisk Astrofysik Center,  Danmarks Grundforskningsfond, and \\
Institut for Fysik og Astronomi, Aarhus Universitet, DK-8000 Aarhus C,
Denmark \and
MICHAEL J. THOMPSON \\
Space \& Atmospheric Physics, The Blackett Laboratory, Imperial College, \\
London SW7 2BZ, UK}

\begin{document}


\setlength{\baselineskip}{15pt}
\dblspace

\author{J{\O}RGEN CHRISTENSEN-DALSGAARD\\
Teoretisk Astrofysik Center,  Danmarks Grundforskningsfond, and \\
Institut for Fysik og Astronomi, Aarhus Universitet, DK-8000 Aarhus C,
Denmark \and
MICHAEL J. THOMPSON \\
Space \& Atmospheric Physics, The Blackett Laboratory, Imperial College, \\
London SW7 2BZ, UK}

\chapter{Rotation of the solar interior}




{\it Helioseismology has allowed us to infer the rotation 
in the greater part of the solar interior with high precision and
resolution.
The results show interesting conflicts with earlier theoretical
expectations, indicating that the the Sun is host to 
complex dynamical phenomena, so far hardly understood.
This has important consequences for our ideas about the
evolution of stellar rotation,
as well as for models for the generation of the solar magnetic field.
Here we provide an overview of our current knowledge about solar
rotation,
much of it obtained from observations from the SOHO spacecraft,
and discuss the broader implications.
}

\section{Introduction}

Solar rotation has been known at least since the early seventeenth
century when, with the newly invented telescope,
Fabricius, Galileo and Scheiner observed 
the motion of sunspots across the solar disk
(for a brief review, see Charbonneau et al.\ 1999).
The rotation of the Sun and other stars originates from 
the contracting interstellar gas clouds from which stars are born;
these clouds share the rotation of the Galaxy.
As the clouds contract, 
they rotate more rapidly, as a result of the conservation of angular
momentum and the reduction in the moment of inertia with contraction
(for a discussion of star formation, see Lada \& Shu 1990).
Although the details of star formation within the contracting clouds
are uncertain and involve
mass loss and interaction with disks around the star which will 
transport angular momentum from one part of the cloud to another,
it is plausible that newly formed star should be spinning quite rapidly.
This is indeed observed: the rotation of the stellar surface
causes a broadening of the lines in the star's spectrum, owing to
the Doppler effect, and from measurements of this effect it is inferred
that many young stars rotate at near the break-up speed, where
the centrifugal force at the equator almost equals gravity.

Stars tend to slow down when they get older. 
At least for stars of roughly solar type, the observations show
that the rotation rate decreases with increasing age
(Skumanich 1972).
This is thought to take place through angular-momentum
loss in a wind, magnetically coupled to the outer parts
of the Sun (e.g.\ Mestel 1968; Mestel \& Spruit 1987). 
The extent to which the slowdown affects the deep interior of
the Sun then depends on the efficiency of the coupling between
the inner and outer parts.
In fact, simple models of the dynamics of the solar interior 
tend to predict that the core of the Sun is rotating
up to fifty times as rapidly as the surface
(e.g.\ Pinsonneault et al.\ 1989).
Such a rapidly rotating solar core could have serious consequences
for the tests of Einstein's theory of general relativity based on
observations of planetary motion: a rapidly rotating core would
flatten the Sun and hence perturb the gravitational field around it.
Even a subtle effect of this nature, difficult to see directly
on the Sun's turbulent surface, might be significant.

\begin{figure}
\begin{center}
\leavevmode\epsfxsize=10cm \epsfbox{\fig/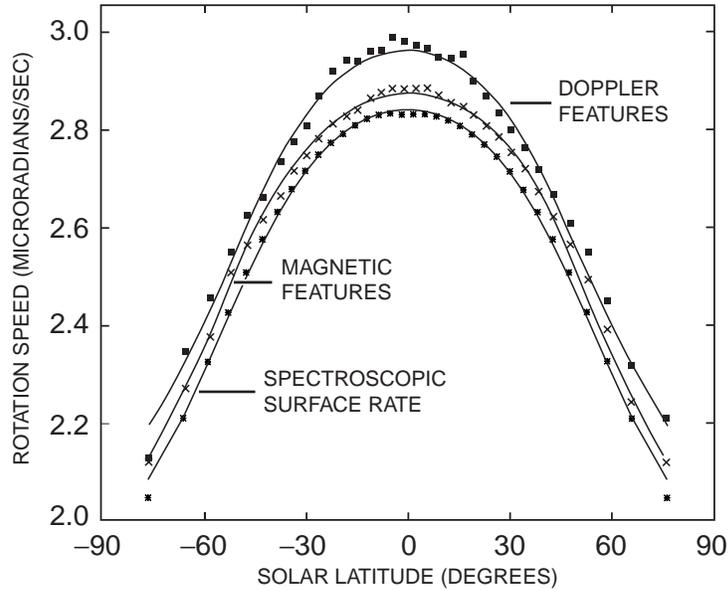}
\end{center}
\caption{
\dblspace
Near-surface solar rotation rates as determined from surface spectroscopic
Doppler-velocity measurements, 
tracking magnetic features, and tracking Doppler features
resulting from large-scale convective flow patterns.
A rotation period
of 36 days corresponds to an angular speed of $2.02$ microrad/s,
and a period of 25 days corresponds to $2.91$ microrad/s.
(Adapted from Snodgrass \& Ulrich\ 1990.) 
}
\label{fig:surf_rot}
\end{figure}

Very detailed observations have been carried out of the solar
surface rotation by tracking the motion of surface features
such as sunspots and, more recently, by Doppler-velocity measurements.
It was firmly established by the nineteenth century, by careful tracking
of sunspots at different latitudes on the Sun's surface,
that the Sun is not rotating as a solid body: at the equator the
rotation period is around 25 days, but it increases gradually
towards the poles where the period is estimated to be in excess of 36 days.
Figure~\ref{fig:surf_rot} shows the near-surface solar rotation determined
from surface Doppler measurements,
as well as from tracking magnetic features and large-scale
convective flow patterns, as a function 
of solar latitude.
The origin of this {\it differential rotation} is almost certainly linked to 
the otherwise dynamic nature of the outer parts of the Sun.
In the outer 29 \% of the Sun's radius (or 200 Mm), energy
is transported by {\it convection}, in rising elements of
warm gas and sinking elements of colder gas: this region is called
the {\it convection zone}
(for an overview of solar structure, see Christensen-Dalsgaard et al.\ 1996).
The convection
can be seen directly using high-resolution observations of the solar surface, 
in the {\it granulation},
with brighter areas of warm gas just arrived at the surface,
surrounded by colder lanes of sinking gas.
The gas motions also transport angular momentum, and hence
provide a link between rotation in different parts of the convection zone.
Furthermore, convection is affected by rotation, which may introduce
anisotropy in the angular momentum transport.
Indeed, it is likely that this transport is responsible for the
differential rotation, although the details are far from understood.

Similarly complex dynamical interactions are also found in the
giant gas\-eous planets (Jupiter, Saturn, Uranus and Neptune)
which, like the Sun, are vigorously convecting as they rotate
(see Ingersoll 1990).
Here the interaction probably gives rise to the banded structures
immediately visible on Jupiter, and more faintly on Saturn. Even
closer to home, the Earth's atmosphere and oceans are
rotating fluid systems and exhibit, among other things, large-scale
circulations and meandering jets such as the jet stream. In all these
systems, rotation plays a significant role in the observed 
dynamical behaviour.

\section{Helioseismic probes of the solar interior}

\begin{figure}
\begin{center}
\leavevmode\epsfxsize=10cm \epsfbox{\fig/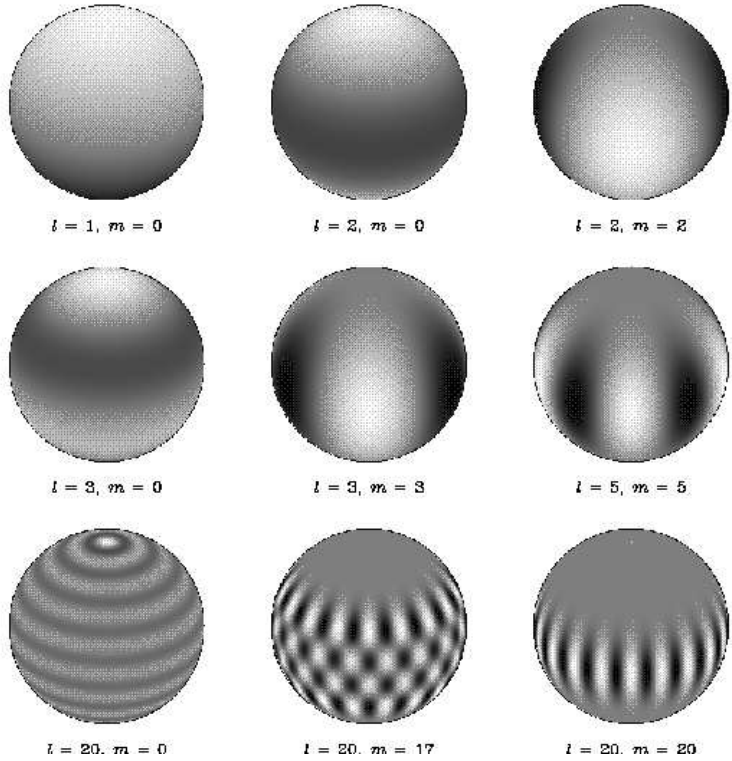} 
\end{center}
\caption{
\dblspace
Examples of spherical harmonic patterns, for different values of the 
degree $l$ and order $m$.
For clarity the polar axis has been inclined $30^\circ$ relative to the page.
}
\label{fig:ylm}
\end{figure}

In recent years, the observation that the Sun is oscillating 
simultaneously in many 
small-amplitude global resonant modes has provided a new diagnostic
of the solar interior. 
The frequencies of these global modes depend on conditions inside the Sun
(Gough \& Toomre 1991; see also Chapter 3, by Chitre \& Antia), 
and so by measuring
these frequencies we are able to make deductions about the state
of the interior. This field is known as {\it helioseismology}. The
observed oscillations are sometimes called five-minute oscillations,
because they have periods in the vicinity of five minutes.
The modes are distinguished not only by their different frequencies, but
also by their different patterns on the surface of the Sun. 
Specifically, the behaviour of a mode is characterized by a 
spherical harmonic 
\begin{equation}
Y_l^m(\theta, \phi) = c_{lm} P_l^m(\cos \theta) \exp(i m \phi)
\end{equation}
as function of co-latitude $\theta$ and longitude $\phi$;
here $P_l^m$ is a Legendre function, and $c_{lm}$ is a normalization constant.
The spherical harmonics are characterized by two 
integer numbers, their {\it degree} $l$ and their {\it azimuthal order} $m$;
a few examples are illustrated in Fig.~\ref{fig:ylm}.
In addition, a mode is characterized by its radial order $n$ which,
approximately, is given as the number of nodes in the radial direction.
The dependence on time $t$ of the oscillation can be written as
$\exp(- i \omega t)$, where $\omega$ is the angular frequency.
Different modes are sensitive 
to different regions of the Sun, depending on their frequency, degree 
and azimuthal order (see also Chapter 3).
In particular, in the radial direction the modes are essentially
confined outside an inner {\it turning point} at the distance
$r = \rt$ from the centre, where $\rt$ satisfies
\begin{equation}
{c(\rt) \over \rt} = {\omega \over {l+1/2}} \; ,
\label{eqn:rt}
\end{equation}
$c$ being the sound speed;
thus low-degree modes extend over much of the Sun while high-degree
modes are confined close to the solar surface.
By exploiting the different sensitivities of the
modes, helioseismology is able to make inferences about localized 
conditions inside the Sun.

\subsection{Rotational effects on the oscillation frequencies}

\label{sec:rot-effects}

One of the factors that affect the mode frequencies is the Sun's
rotation.
The dominant effect of rotation on the oscillation frequencies is
quite simple: the oscillation patterns illustrated in Fig.~\ref{fig:ylm}
actually correspond to waves running around the equator.
Patterns travelling in the same direction as the rotation of the
Sun would appear to rotate a little faster, patterns rotating
in the opposite direction a little more slowly.
When observed at a given position on the Sun the oscillations in
the former case would have slightly higher frequency, and in the
latter case slightly lower frequency, than if the Sun had not been rotating.
The frequency difference between these two cases therefore provides
a measure of the rotation rate of the Sun.

\begin{figure}[!ht]
\begin{center}
\leavevmode\epsfxsize=6cm\epsfbox{\fig/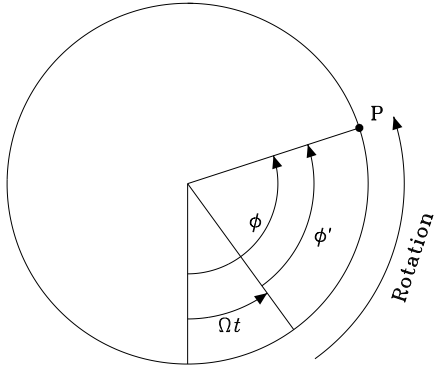}
\end{center}
\caption{
\dblspace
\label{fig:rotgeom}
Geometry of rotational splitting, in a star rotating with
angular velocity $\Omega$.
The point {\bf P} has
longitude $\phi'$ in the system rotating with the star
and longitude $\phi = \phi' + \Omega t$ in the inertial system.
}
\end{figure}

This simple description can be made more precise by noting that,
as a function of longitude and time $t$, the oscillations behave
as $\cos(\omega t - m \phi)$ (apart from an arbitrary phase).
We now consider a star that is rotating uniformly,
with angular velocity $\Omega$, and introduce
a coordinate system rotating with the star, with longitude $\phi '$;
the longitude $\phi$ in an inertial frame is related to $\phi'$
by $\phi' = \phi - \Omega t$ (cf.\ Fig.~\ref{fig:rotgeom}).
We furthermore consider an oscillation with frequency $\omega_0$ in
the rotating frame, and hence varying as $\cos(m \phi' - \omega_0 t)$.
Consequently, the oscillation as observed from the inertial frame
depends on $\phi$ and $t$ as
$$
\cos(m \phi - m \Omega t - \omega_0 t) \equiv \cos ( m \phi -\omega_m t) \; ,
$$
where
\begin{equation}
\omega_m = \omega_0 + m \Omega \; .
\label{simple-split}
\end{equation}
Thus the frequencies are split according to $m$,
the separation between adjacent values of $m$ being simply
the angular velocity;
this is obviously just the result of the advection of the
wave pattern with rotation.

In reality other effects must be taken into account to describe
the frequency shifts caused by rotation,
which are often referred to as the {\it rotational frequency splitting}.
The {\it Coriolis force} affects the dynamics of the oscillations
and hence their frequencies, although it turns out that for the
modes observed in the five-minute region this effect is modest;
owing to the slow rotation of the Sun centrifugal effects are
essentially negligible and in any case have a different functional 
dependence on $m$ (see Section 5.2.2).
However, the variation of angular velocity $\Omega(r, \theta)$
with position in the Sun must be taken into account%
\footnote{Specifically, $\Omega(r, \theta)$ refers to the 
azimuthal component of the azimuthally
averaged flow field within the Sun.}.
Each mode feels an average angular velocity,
where the average is determined by the mode's frequency, degree
and azimuthal order
(Hansen, Cox \& van Horn 1977; Gough 1981).
The precise form of this spatial average is described
by a weight function, or {\it kernel}, such that 
\begin{equation}
\omega_{nlm} = \omega_{nl0} +
m \int_0^R \int_0^\pi K_{nlm}(r,\theta) \Omega(r, \theta)
r \dd  r \dd \theta \; ,
\label{rot-split}
\end{equation}
where $R$ is the total radius of the Sun;
the kernels $K_{nlm}$ can be calculated from the eigenfunctions
for a non-rotating model (Schou, Christensen-Dalsgaard \& Thompson 1994).
It might be noted that
the kernels depend only on $m^2$, so that the rotational splitting
$\omega_{nlm} - \omega_{nl0}$ is an odd function of $m$.
Also, the kernels are symmetrical around the equator;
as a result, the rotational splitting is only sensitive to the
component of $\Omega$ which is similarly symmetrical.
In the special case where $\Omega = \Omega(r)$ is independent of $\theta$,
the integral in \Eq{rot-split} is independent of $m$ and hence the
rotational splitting is simply proportional to $m$;
note that this is the same linear dependence on $m$ as
for the simple effect of advection (cf.\ Eq.\ \ref{simple-split}).

These weight functions vary from mode to mode.
As already indicated in Fig.~\ref{fig:ylm}, modes with $m = l$ are concentrated
near the equator, increasingly so with increasing $l$, whereas modes
of lower azimuthal order extend to higher latitudes. Thus modes with
$m=l$ feel an average of the rotation near the equatorial
plane, whereas modes of lower azimuthal order sense the average 
rotation over a wider range of latitudes. In a similar manner, 
the high-degree five-minute modes 
(i.e., with large values of $l$) sense only conditions near the surface
of the Sun, while modes of low degree feel conditions averaged over
much of the solar interior (cf.\ Eq.\ \ref{eqn:rt}).

\begin{figure}
\begin{center}
\leavevmode\epsfxsize=12cm \epsfbox{\fig/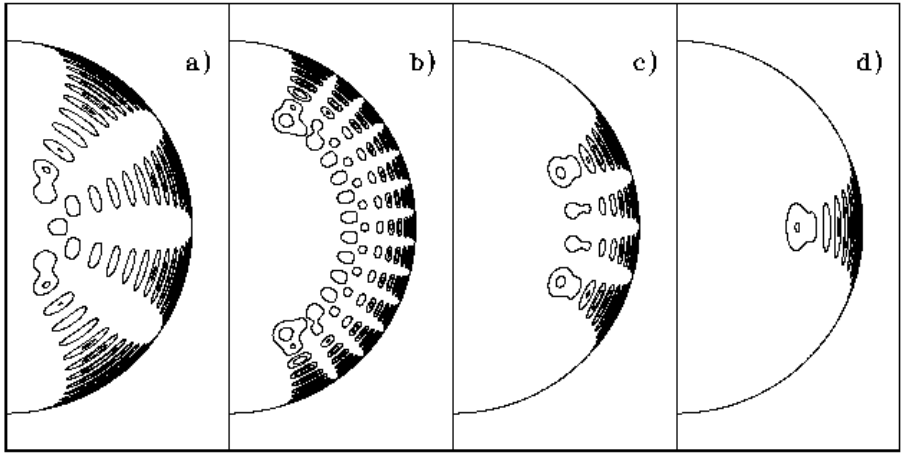} 
\end{center}
\caption{
\dblspace
Contour plot of kernel
weight functions determining the sensitivity of different modes to the
solar internal rotation.
All modes have frequencies near 2~mHz;
their degree $l$ and azimuthal order $m$ are, from left to right:
$(l,m) = (5,2)$, $(20,8)$, $(20,17)$, and $(20,20)$. Note that the 
kernels in panels (a) and (b) have roughly the same latitudinal
extent, because $m/(l+1/2)$ is almost
the same for these modes, but the lower-$l$
mode senses the deeper regions of the Sun; also that for fixed $l$
the modes are more confined to low latitudes as $m$ increases.
}
\label{fig:kernels}
\end{figure}

These properties can be illustrated by a few examples of weight functions,
as shown in Fig.~\ref{fig:kernels}.
The observed modes include some that penetrate essentially to the
solar centre, others that are trapped very near the surface, and
the whole range of intermediate penetration depths, with a similar
variation in latitudinal extent. 
Thus the observed frequency splittings provide a similarly wide range
of averages of the internal rotation.

\begin{figure}[!ht]
\begin{center}
\leavevmode\epsfxsize=9cm\epsfbox{\fig/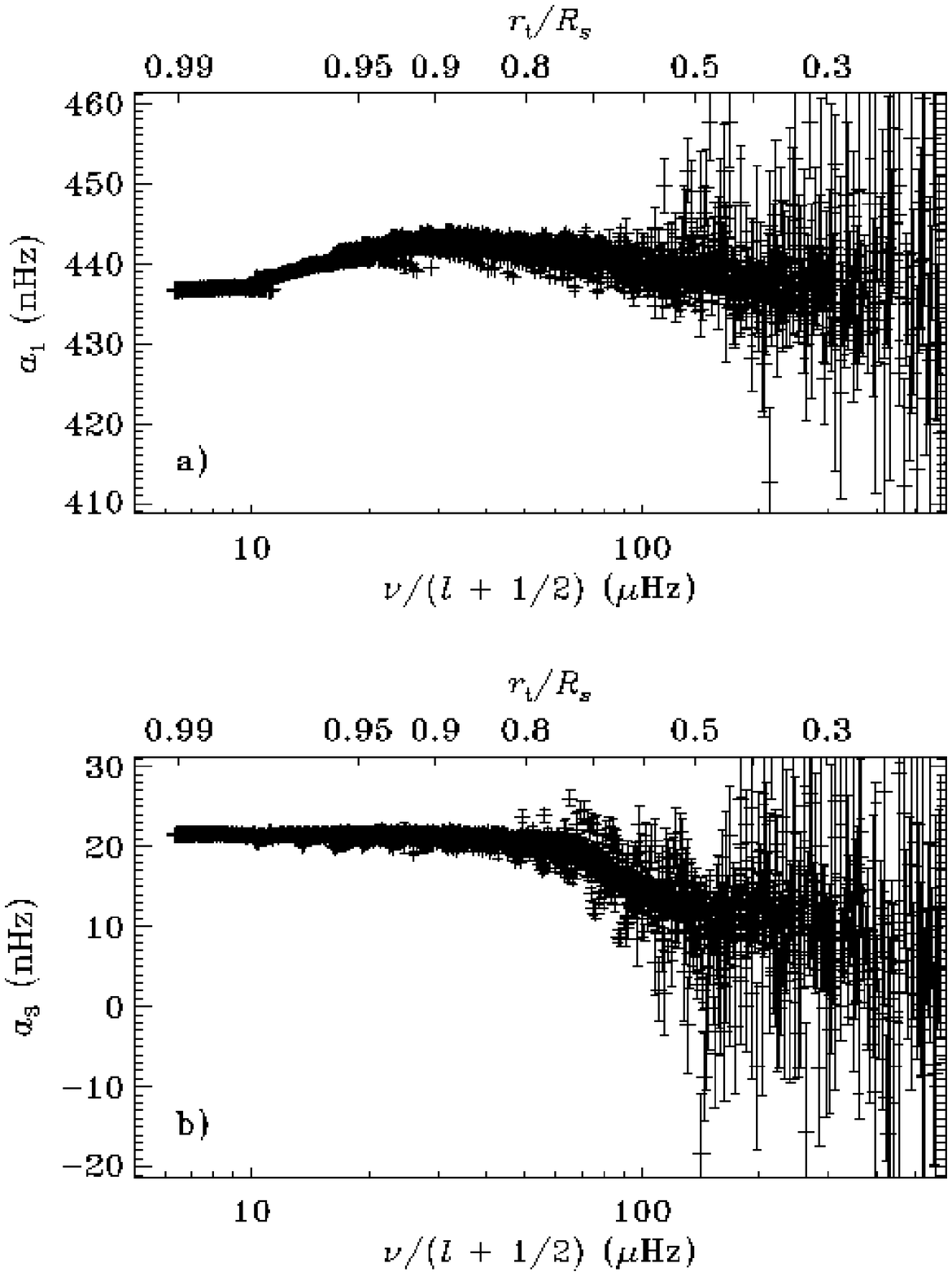}
\end{center}
\caption{
\dblspace
\label{fig:acoeff}
Observed $a$ coefficients $a_1$ and $a_3$ from the expansion 
given in \Eq{acoeff} of frequency splittings from SOI-MDI 
observations.
The error bars correspond to one standard deviation.
The upper abscissas indicate the turning-point radius,
defined by \Eq{eqn:rt}.
}
\end{figure}

\subsection{Data on rotational splitting}

The rotational splittings must of course be measured from precise observations
of the Sun's global oscillations. Several instruments and observational 
networks are engaged in this work (see Section 5.3). For practical 
considerations (including the effects of mode realization arising from 
the finite lifetime of the modes), the observers commonly do not measure the
individual rotational splittings for each $m$ value but instead fit to the 
observational data an expression of the form
\begin{equation}
\nu_{nlm}
= \nu_{nl0}
+ \sum_{j=1}^{j_{\rm max}} a_j (n,l) \, \PP_j^{(l)} {(m)} \; .
\label{acoeff}
\end{equation}
Here then the inferred quantities which convey information about the
rotation are the so-called $a$ coefficients $a_j(n,l)$, and the 
$\PP_j^{(l)}$ are polynomials of degree $j$,
suitably chosen to obtained statistically independent
coefficients (cf.\ Schou et al.\ 1994). The odd $a$ coefficients (i.e., odd
values of $j$) contain 
information about the rotation through the advection and Coriolis effects
on the modes, and these are used to infer the Sun's internal rotation. 

Figure \ref{fig:acoeff} shows the first two odd $a$ coefficients obtained
from observations with the SOI-MDI experiment on the SOHO spacecraft.
To provide some indication of the variation of rotation with depth,
the data are shown against $\nu/(l+1/2)$ ($\nu = \omega/2 \pi$
being the cyclic frequency), which according to \Eq{eqn:rt}
determines the location of the turning point, $r = \rt$, of the mode.
The turning-point radius is indicated as the upper abscissas.
The increase in $a_1$ with increasing $\nu/(l+1/2)$, i.e.,
with increasing depth of penetration of the modes,
indicates that the rotation rate increases with depth in the outer
5 -- 10 \% of the solar radius.
Also, it should be noticed that $a_3$ decreases towards zero for those
modes that penetrate below $r = 0.8 R$, as is indeed the case for the
higher $a$ coefficients also;
thus the rotational splitting tends to become linear in $m$,
suggesting that rotation tends towards latitude independence 
in the inner parts of the Sun (cf.\ Section~\ref{sec:rot-effects}).
These crude inferences are confirmed by the more detailed analysis
presented in Section~\ref{sec:solrot}.

The even $a$ coefficients contain information about aspects of
the Sun which affect modes of azimuthal orders $\pm m$ equally: these 
include centrifugal distortion and other departures of the Sun from 
spherical symmetry. The even coefficients are generally small in value and
are not used for determining the form of the rotation, though they 
can be used to make inferences about the Sun's internal asphericity;
this is discussed in Chapter 3 of the present volume.

\subsection{Inversion for solar internal rotation}

The wealth of data on rotational splitting allows the determination of the
detailed variation of rotation with position in the solar interior.
Modes of high degree, trapped near the surface, provide
measures of the rotation of the superficial layers of the Sun.
Having determined that, its effect on the somewhat more deeply
penetrating modes can be eliminated, leaving just a measure
of rotation at slightly greater depths.
In this way, information about rotation in
the Sun can be `peeled' layer by layer, much
as one could an onion, in a way that allows us to obtain a complete
image of solar internal rotation.

It is fairly evident that this process gets harder, the deeper
one attempts to probe, since fewer and fewer modes penetrate to
the required depth;
furthermore, the effect of rotation decreases because of the smaller
size of the region involved.
Thus the rotation of the solar core is difficult to determine.
Similarly, all modes are affected by the
equatorial rotation while only modes of low $m$ extend to the vicinity 
of the poles, and the polar regions have relatively little effect
on the oscillations, complicating the
determination of the high-latitude rotation.
However, as we shall see the quality of current data is such that
the angular velocity can be determined quite near the poles,
at least in the outer parts of the convection zone.

The analysis of the oscillation data to infer properties
of the solar interior is often characterized as
{\it inversion} (Gough \& Thompson 1991).
To illustrate aspects of these procedures, we write \Eq{rot-split} as
\begin{equation}
\Delta_i = \int K_i(\bx) \Omega(\bx) \dd \bx \; ,
\label{inv-prob}
\end{equation}
where the label $i$ stands for the modes $(n, l, m)$ and $\bx$
replaces the coordinates $(r, \theta)$; 
also, $\Delta_i = m^{-1}(\omega_{nlm} - \omega_{nl0})$
are the observed data. Likewise, kernels similar to the $K_i(\bx)$ can be
derived if the data $\Delta_i$ are in the form of $a$ coefficients, as
in \Eq{acoeff}.
The goal of the analysis is obviously to infer the properties
of $\Omega(\bx)$ from the data, taking into account also the
observational errors.
In most cases considered so far, the analysis corresponds either
implicitly or explicitly to making linear combinations of the data;
thus to infer $\Omega$ at a point $\bx_0 = (r_0, \theta_0)$, say,
coefficients $c_i(\bx_0)$ are determined such that, as far as possible,
\begin{equation}
\Omega(\bx_0) \simeq \bar \Omega(\bx_0) = \sum_i c_i(\bx_0) \Delta_i \; ,
\label{inv-sol1}
\end{equation}
where $\bar \Omega$ is the inferred approximation to the true angular velocity.
From \Eq{inv-prob} it then follows that
\begin{equation}
\bar \Omega(\bx_0) = \int \CK(\bx_0, \bx) \Omega(\bx) \, \dd \bx \; ,
\label{inv-sol2}
\end{equation}
where the {\it averaging kernels} $\CK$ are given by
\begin{equation}
\CK(\bx_0, \bx) = \sum_i c_i(\bx_0) K_i(\bx) \; .
\label{av-ker}
\end{equation}
Also, if the standard errors $\sigma(\Delta_i)$ of the observations are
known, the error in the inferred angular velocity 
$\bar \Omega(\bx_0)$ can be found from \Eq{inv-sol1}.

It is evident from \Eq{inv-sol2} that the coefficients $c_i(\bx_0)$
should be determined such that $\CK(\bx_0, \bx)$ is as far as possible
localized near $\bx = \bx_0$.
The extent of $\CK$ provides a measure of the resolution of the inversion.
At the same time, it must be ensured that the error on $\bar \Omega(\bx_0)$
is reasonable.
In general, there is a trade-off between error and resolution, determined
by one or more parameters of the procedure.

In one commonly used inversion procedure, the so-called
regularized least-squares (or RLS) procedure, the solution
is parametrized and the parameters are determined through
a least-squares fit of the data to the right-hand side of \Eq{inv-prob}.
To ensure that the solution is well-behaved and the errors are of
reasonable magnitude, the fit is regularized by limiting also,
for example, a measure of the square of the second derivative of
the solution, thus penalizing rapid variations.
From the results of the fit the coefficients $c_i(\bx_0)$, and
hence the averaging kernels, can be determined.
In a second class of procedures, the optimally-localized averages
(or OLA) procedures, the $c_i(\bx_0)$ are determined such as to
obtain a localized averaging kernel $\CK(\bx_0, \bx)$,
while at the same time limiting the errors.
Details of these procedures and their results were
given by Schou et al.\ (1998).

As already remarked, the weight functions (Fig.~\ref{fig:kernels})
are symmetrical around the
solar equator, and so we can infer only the 
similarly symmetric component of rotation.
This must be kept in mind in the following, when interpreting the results.
We note that this restriction can be avoided by applying {\it local
helioseismology} techniques to the data: such techniques
are described elsewhere in this volume by Kosovichev.

\section{The solar internal rotation}

\label{sec:solrot}

\begin{figure}
\begin{center}
\leavevmode\epsfxsize=9cm\epsfbox{\fig/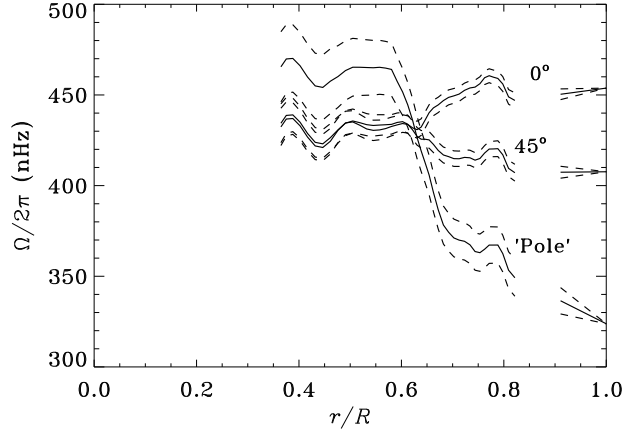}
\end{center}
\caption{
\dblspace
\label{fig:tenerife-rot}
Rotation rate $\Omega/2 \pi$ inferred from inversion of 
$a$ coefficients from BBSO (Libbrecht 1989),
targeted at the equator, latitude $45^\circ$ and the pole.
The dashed lines indicate the $1-\sigma$ error limits.
The inversion was only possible between $0.4 R$ and $0.8 R$;
for clarity, the results have been connected with the directly
observed surface rates.
(Adapted from Christensen-Dalsgaard \& Schou 1988.)
} 
\end{figure}

Early helioseismic data on rotational splittings provided information
only about the modes with $m \simeq \pm l$; as a result,
they were sensitive mainly to rotation near the equator.
Duvall et al.\ (1984) showed that there was relatively little variation
of rotation with depth; in particular, there were no significant
indications of a rapidly rotating core.
A few years later initial data on the dependence of the splitting on $m$
were obtained (Brown 1985; Libbrecht 1988, 1989).
Strikingly, they indicated that the surface latitudinal differential rotation 
persisted through the convection zone, whereas there was little indication
of variation with latitude in the rotation beneath the convection zone
(e.g.\ Brown \& Morrow 1987; Brown et al.\ 1989;
Christensen-Dalsgaard \& Schou 1988; see also Fig.~\ref{fig:tenerife-rot}).

\begin{figure}
\begin{center}
\leavevmode\epsfxsize=8truecm \epsfbox{\fig/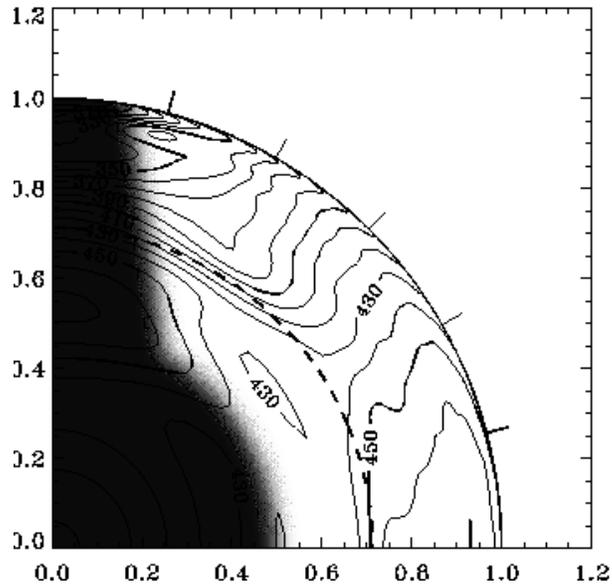}
\end{center}
\caption{
\dblspace
Rotation of the solar envelope inferred from observations by
SOI-MDI (Schou et al.\ 1998).
The equator is at the horizontal axis and the pole is at the vertical
axis, both axes being labelled by fractional radius.
Some contours are labelled in nHz, and, for clarity, selected
contours are shown by bolder curves.
The dashed circle indicates the base of
the convection zone and the tick marks at the edge of the outer
circle are at latitudes $15^\circ$, $30^\circ$, $45^\circ$, $60^\circ$, and
$75^\circ$.
The shaded area indicates the region in the Sun where no reliable inference
can be made with the current data.
}
\label{fig:fig3}
\end{figure}

\begin{figure}
\begin{center}
\leavevmode\epsfxsize=10truecm \epsfbox{\fig/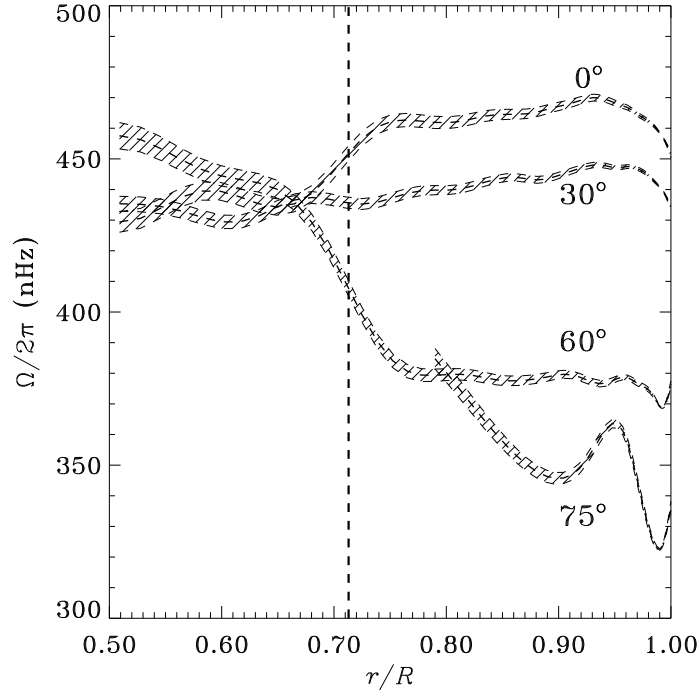}
\end{center}
\caption{
\dblspace
Rotation of the solar envelope as a function of radius,
at the latitudes indicated, inferred from observations by
SOI-MDI (Schou et al.\ 1998).
The heavy dashed line marks the base of the convection zone.
}
\label{fig:fig3q}
\end{figure}

In the last few years the amount and quality of helioseismic data on
solar rotation have increased dramatically, as a result of several
ground-based and space-based experiments (Duvall 1995).
The LOWL instrument of the High Altitude Observatory has provided
high-quality data on modes of low and intermediate degree over
the past more than five years.
The BiSON and IRIS networks, observing low-degree modes in Doppler
velocity integrated over the solar disk, have yielded increasingly
tight constraints on the rotation of the solar core, while the GONG
six-station network
is setting a new standard for ground-based helioseismology
(Harvey et al.\ 1996).  Further, the SOI-MDI experiment on SOHO 
(Scherrer et al.\ 1995) has yielded
a wealth of data on modes of degree up to 300, allowing 
a detailed analysis of the properties of rotation in the
convection zone. The results we present below are the combined knowledge
that has emerged from these observational efforts
(e.g.\ Tomczyk, Schou \& Thompson 1995; Thompson et al.\ 1996; 
Corbard et al.\ 1997, 1999; Di Mauro \& Dziembowski 1998; Schou et al.\ 1998).
Some of these results are illustrated in Figs~\ref{fig:fig3} and 
\ref{fig:fig3q}. These are the result of a so-called regularized least
squares inversion applied to SOI-MDI data 
(Schou et al.\ 1998). Figure~\ref{fig:fig3}
shows the inferred rotation as a contour plot, in the region where the
results are judged to be reliable; Fig.~\ref{fig:fig3q} shows the radial
cuts through the same solution, at a few selected latitudes.

\subsection{Rotation of the solar convection zone}

In discussing what we now know about the rotation inside the Sun, we
shall start from the near-surface layers and work towards the centre. 
As we have already discussed, the outer 29 per cent of the Sun 
is convectively unstable. Before helioseismology,
models predicted that the rotation inside the convection zone would 
organize itself on cylinders aligned with the rotation axis
(Glatzmaier 1985; Gilman \& Miller 1986; see also Section~\ref{sec:models}).
Thus the rotation at depth at, say, equatorial latitudes would match the  
surface rotation at high latitudes, rather than the faster equatorial rotation 
at the surface, and so at given latitude the rotation in the convection
zone would decrease with depth. Helioseismology has shown that this is not 
so: to a first approximation it is more accurate to say that the rotation at 
a given latitude is nearly constant with depth, or to put it another way
the differential rotation seen at the surface imprints itself through the
convection zone. This finding is clearly visible in Figs~\ref{fig:fig3}
and \ref{fig:fig3q}.
In detail, the situation is more complicated. 
At low latitudes, immediately beneath the solar surface the rotation 
rate actually initially {\it increases} with depth. The equatorial rotation
reaches a maximum at a depth of about 50 Mm 
(i.e., about 7 per cent of the
way in from the surface to the centre of the Sun): at this point, the
rotation rate is about 5 per cent higher than it is at the surface.
This is consistent with a variety of surface measurements of rotation.
Tracking sunspots tends to give a slightly
higher rotation rate than that obtained by making
direct spectroscopic measurements of the velocity of the surface
(Korzennik et al.\ 1990; see also Fig.~\ref{fig:surf_rot}).
Probably the reason is that 
the sunspots extend to some depth below the surface, and so are 
dragged along at a rate that is more similar to the subsurface rotation
a few per cent beneath the surface which helioseismology has revealed.

\begin{figure}
\begin{center}
\leavevmode\epsfxsize=12truecm \epsfbox{\fig/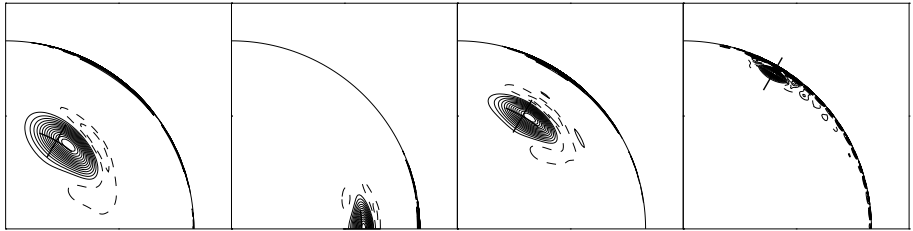}
\end{center}
\caption{
\dblspace
Averaging kernels for the inversion solution shown in Fig.~\ref{fig:fig3},
at selected radii and latitudes in the Sun, which are (from left to right) as 
follows: $0.540R$, $60^\circ$; 
$0.692R$, $0^\circ$; $0.692R$, $60^\circ$; $0.952R$, $60^\circ$. 
The corresponding locations are indicated with crosses.
Positive contours are shows as solid lines, negative contours as
dashed lines. (Adapted from Schou et al.\ 1998.)
}
\label{fig:avkers}
\end{figure}

Figure~\ref{fig:avkers} shows averaging kernels (cf. Eq.~\ref{av-ker})
corresponding to the 
solution in Fig.~\ref{fig:fig3}, at a few locations inside the Sun. 
These give an indication of the resolution attained by the inversion. Note
that the radial resolution is finer for points high in the convection zone
and poorer in the radiative interior. This is largely a reflection of the
local vertical wavenumber of the waves at the various depths. 

\begin{figure}
\begin{center}
\leavevmode\epsfxsize=12truecm \epsfbox{\fig/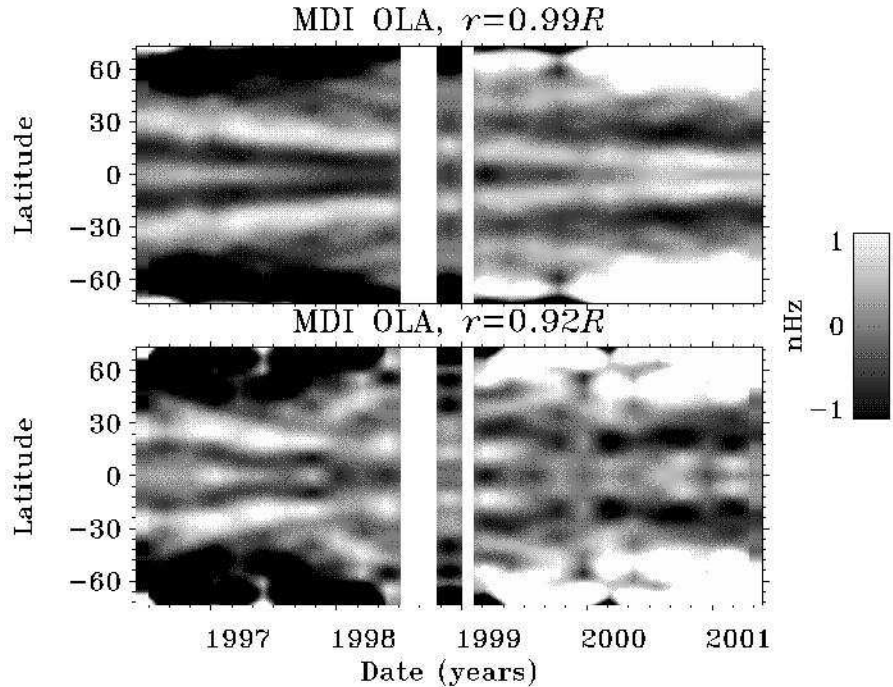}
\end{center}
\caption{
\dblspace
The evolution with time of the fine structure in the near-surface
solar rotation, based on observations from the SOI-MDI
instrument on the SOHO spacecraft, after subtraction of 
the time-averaged rotation rate.
The result is represented as a function of time (horizontal axis)
and latitude (vertical axis), the grey scale at the right giving
the scale in nHz;
1 nHz corresponds to a speed of around 4 m/s at the equator.
The banded structure, apparently converging towards the equator as time
goes by, should be noted. 
The vertical white bands correspond to time intervals when the
spacecraft was temporarily inactive.
(Adapted from Howe et al.\ 2001.)
}
\label{fig:zonal}
\end{figure}

The rotational velocity at the surface of the Sun is about 2 km per second,
dropping off rather smoothly towards higher latitudes.
However, it has now been found that superimposed on this are bands of
faster and slower rotation, a few metres per second higher or lower than
the mean flow (Kosovichev \& Schou 1997; Schou 1999).
The origin of this behaviour, illustrated in Fig.~\ref{fig:zonal},
is not understood, but it is
reminiscent of the more pronounced banded flow patterns seen on 
Jupiter and Saturn. 
Evidence for such bands had been obtained previously from
direct Doppler measurements on the solar surface (Howard \& LaBonte 1980).
However, the seismic inferences have shown that they extend
to a depth probably exceeding 40 Mm beneath the surface
(Howe et al.\ 2000a).
Moreover, these bands migrate from high latitudes
towards the equator over the solar cycle.

It has been customary to represent the directly measured surface
rotation rate in terms of a simple low-order expansion in
$\sin \psi$, where $\psi$ is latitude on the Sun. 
This in fact quite successfully captured the observed behaviour;
however, since the solar rotation axis is close to the plane of
the sky, direct measurements of rotation near the poles are
difficult and uncertain.
Strikingly, the helioseismic results have shown a marked departure
from this behaviour, at latitudes above about $60^\circ$:
Relative to the simple fit, the actual rotation rate decreases
quite markedly there.
The origin or significance of this behaviour is not yet understood.
There is also evidence, hinted at in Fig.~\ref{fig:fig3}, of
a more complex behaviour of rotation at high latitudes.
Some analyses have shown a `jet', i.e., a localized region of more
rapid rotation, at a latitude around $75^\circ$ and a depth of about 
35 Mm beneath the solar surface
(Schou et al.\ 1998).
Also, evidence has been found that the rotation rate shows substantial
variations in time at high latitudes, over time scales of order months.
It is probably fair to say that the significance of these results is still
somewhat uncertain, however.
Also, it should be kept in mind, as mentioned above, that
the results provide an average of rotation in the Northern and
Southern hemispheres and, evidently, an average over the observing
period of at least 2 -- 3 months.
Thus the interpretation of the inferred rotation rates in terms
of the actual dynamics of the solar convection zone is 
not straightforward.

\subsection{The tachocline}

At the base of the convection zone, a remarkable transition occurs: the
variation of rotation rate with latitude disappears, so that the 
region beneath the convection zone rotates essentially rigidly, at
a rate corresponding to the surface rate at mid-latitudes
(see Fig.~\ref{fig:fig5}). The region
over which the transition occurs is very narrow, no more than a few
per cent of the total radius of the Sun (e.g.\ Kosovichev 1996,
Charbonneau et al. 1999). This layer has been 
called the {\it tachocline} (Spiegel \& Zahn 1992).
Why the differential rotation
does not persist beneath the convection zone is not yet known, but it
is possible that a large-scale weak magnetic field permeates the inner
region and enforces nearly rigid rotation by dragging the gas along at
a common rate (Gough \& McIntyre 1998).
Such a field is quite possible as a relic from the original
collapsing gas cloud from which the Sun condensed. 

The discovery of the tachocline, and of the form of the rotation in 
the convection zone, has led to an adjustment of our theories of the
{\it solar dynamo} (see the chapter by Choudhuri in this volume).
One idea is that the dynamo 
action consists of two components: a twisting of the magnetic 
field by the motion of convective elements, and a shearing out
of the field by differential rotation. 
Prior to the helioseismic findings, the simulations of rotation
implied that the radial gradient of differential rotation in the 
convection zone could provide the second ingredient, so it was
thought that the dynamo action occurred in that region. Now, however,
the tachocline with its very substantial radial gradient seems
a more likely location for the dynamo action producing the large-scale
magnetic field (Gilman, Morrow \& DeLuca 1989).

\begin{figure}
\begin{center}
\leavevmode\epsfxsize=12truecm \epsfbox{\fig/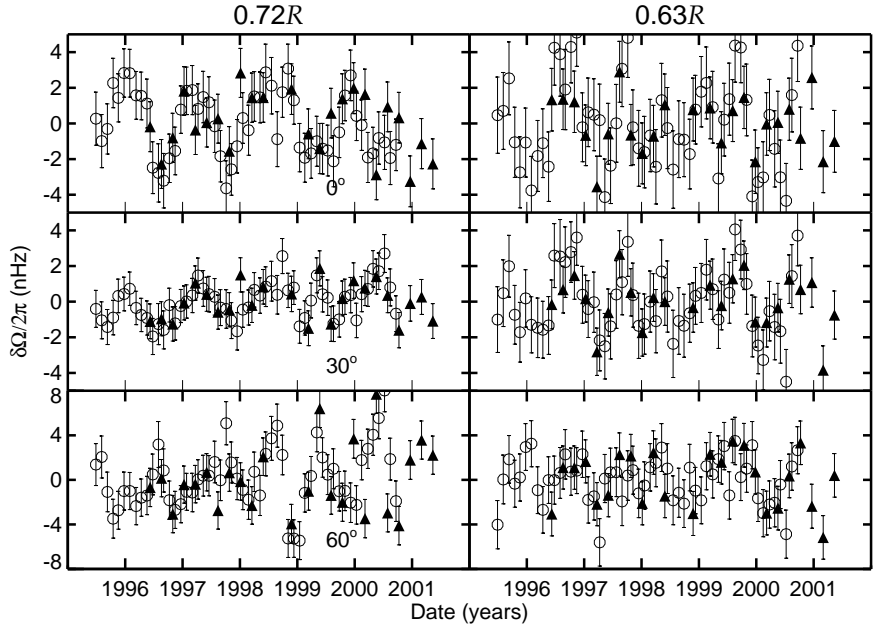}
\end{center}
\caption{
\dblspace
Deviation from the mean rotation rate inferred from inversions
at various locations near the base of the convection zone, as a
function of time.
Open circles represent results from the GONG network, and filled
triangles are from the SOI-MDI experiment.
(From Howe et al.\ 2001.)
}
\label{fig:tachoosc}
\end{figure}

If the magnetic field is built up in the tachocline over the course of
the solar cycle, one might expect to see variations over time in the
tachocline properties, including perhaps the rotation rate there. 
Variations in the rotation rate have in fact been found in the 
vicinity of the tachocline, but the timescale was unexpected. Rather than
the 11-year timescale of the solar cycle, Howe et al. (2000b) have reported
quasi-periodic variations in the equatorial region just above and below the 
tachocline with a period of 1.3 years (Fig.~\ref{fig:tachoosc}). The 
oscillations are revealed by subtracting out the temporal mean rotation
rate at each location and looking at the residuals as a function of time.
The equatorial variations at radius $0.72R$ show the 1.3-year oscillation
most strongly, and the variations at radius $0.63R$
exhibit the same period but are in antiphase. This 
implies that the radiative interior also partakes in the temporal variation,
and suggests perhaps a back-and-forth exchange of angular momentum between
the two locations. The signal is much weaker at $30^\circ$ latitude,
and the variations at higher latitudes are possibly not significant. These
findings are exciting because they may represent the first direct observation
of variability at the seat of the solar dynamo.

\begin{figure}
\begin{center}
\leavevmode\epsfxsize=10cm \epsfbox{\fig/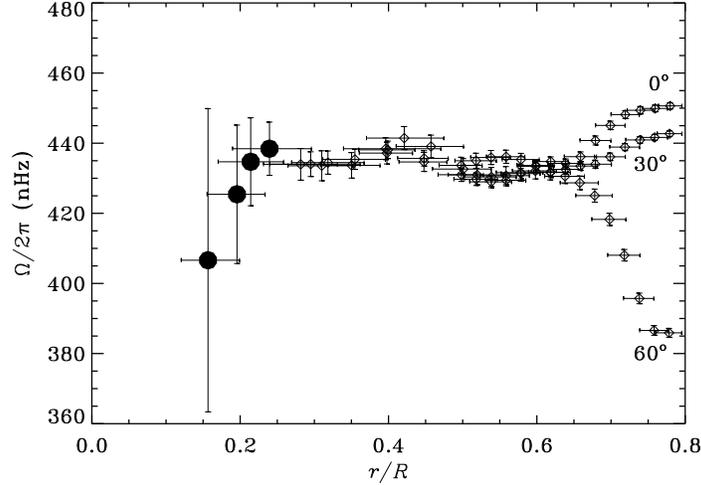}
\end{center}
\caption{
\dblspace
The inferred rotation as a function of fractional radius inside the Sun at
three solar latitudes: the equator, $30^\circ$ and $60^\circ$;
the vertical axis shows the rotation frequency in nHz.
The vertical error bars 
indicate the statistical uncertainty on the rotation rate ($\pm 1$
standard deviation),
wheras the horizontal bars provide a measure of the radial resolution
of the inversion.
Note that the result becomes much more uncertain in the deep interior,
where furthermore the different latitudes cannot be separated.
The observations used to infer the rotation were from the LOWL instrument
and the BiSON network (Chaplin et al.\ 1999).
}
\label{fig:fig5}
\end{figure}

\subsection{The radiative interior}

Even deeper in the Sun, 
in the radiative interior,
the helioseismic results on the rotation 
are more uncertain due to the fact that so few of the observed 
five-minute modes (only the low-degree modes) 
have any sensitivity to this region.
Indeed, the results have been somewhat contradictory
(for a review, see Eff-Darwich \& Korzennik 1998), some indicating 
rotation faster than the surface rate and others indicating rotation
slower than or comparable to the rotation rate at the base of the convection
zone;
an example is illustrated in Fig.~\ref{fig:fig5}.
However, down to within 15 per cent of the solar radius from the 
centre, which is the deepest point at which present observations 
permit localized inferences to be made, all the modern results agree that
the rotation rate is not more than a factor two different from the 
surface rate: thus early models which predicted that the whole of
the nuclear-burning core was rotating much faster are firmly ruled out. 
Again, this finding
would be consistent with a magnetic field linking the core
to the bulk of the radiative interior.

\section{Modelling solar rotation}

\label{sec:models}

Although helioseismology has provided us with a remarkably detailed
view of solar internal rotation, the theoretical understanding
of the inferred behaviour is still incomplete.
In the convection zone, the problem is to model the complex
combined dynamics of rotation and convection, the latter occurring
on scales from probably less than a few hundred kilometres 
to the scale of the entire convection zone and time scales from minutes
to years. Viscous dissipation is estimated to occur on even smaller
spatial scales, of order 0.1~km or less.
Capturing this range of scales is entirely outside the possibility
of current numerical simulations; thus simplifications are required.
Detailed simulations of near-surface convection, on a scale of a few Mm,
have been remarkably successful in reproducing the observed properties
of the granulation (Stein \& Nordlund 1989, 1998)
but are evidently not directly relevant to
the question of rotation.
Simulations of the entire convection zone are necessarily
restricted to rather large scales and hence cannot capture the
near-surface details.
Such simulations, therefore, typically exclude
the outer 30 Mm of the convection zone.
Early examples of such simulations by Gilman and Glatzmaier,
of fairly limited resolution, showed a tendency for rotation to
organize itself on cylinders
(Glatzmaier 1985; Gilman \& Miller 1986): the rotation rate depended primarily
on the distance to the rotation axis. 
The convection itself in these simulations was
dominated by so-called {\it banana cells} -- long, thin, 
large-scale convection cells oriented in the north-south direction --
as seen also, for example, in laboratory convection as observed in
the GFFC Spacelab experiment (Hart et al. 1985).

Rotation on cylinders is
predicted for simple systems by the
{\it Taylor-Proudman theorem}, which may be derived as follows. The
velocity $\bu$ in an inviscid fluid in a gravitational potential $\Phi$
satisfies an equation of motion
\begin{equation}
{\partial\bu\over\partial t} + \bu\cdot\nabla\bu = 
-{1\over\rho}\nabla p - \nabla\Phi\;,
\label{momentum}
\end{equation}
where $p$ and $\rho$ are pressure and density respectively, and $t$
is time. Suppose that the velocity is purely rotational and independent
of time, so in cylindrical coordinates $(s,\phi,z)$ we write
$\bu = s\Omega(s,z)\be_\phi$, where $s$ is the distance from the 
rotation axis, $z$ distance along the axis, and $\be_\phi$ is a 
unit vector in the azimuthal direction. Taking the curl of \Eq{momentum} gives
\begin{equation}
-s{\partial\,\Omega^2\over\partial z}\be_\phi
= {1\over\rho^2}\nabla\rho\times\nabla p\;.
\label{TaylorProudman}
\end{equation}
In the solar convection zone the fluid is essentially isentropic (uniform
specific entropy) and so the pressure can be regarded as a function of
the density alone: hence the right-hand side of \Eq{TaylorProudman} is
zero, and thus $\Omega$ depends only on $s$, not on $z$, i.e., the rotation
rate is uniform on cylindrical surfaces. 

If the Taylor-Proudman theorem applied in the Sun,
the rotation rate on a given cylinder would obviously be
observable where the cylinder intersected the surface;
thus the observed decrease of rotation rate with increasing latitude
would correspond to a similar decrease of rotation rate with depth at,
say, the equator. 
The actual solar rotation, shown in Fig.~\ref{fig:fig3}, is obviously
very different from this behaviour and from
the simulation results mentioned above.
The overall variation of rotation within the convection zone
is evidently predominantly with latitude, with little variation in the radial
direction except in the tachocline.
Given the necessary simplifications of the calculations, their
failure to model solar rotation is
perhaps not surprising.
In particular, the effects of smaller-scale turbulence
(beneath the smallest scale resolved in the simulation)
is typically represented as some form of viscosity;
it was suggested by Gough (1976), and later by others, that the
effect of rotation on the small-scale motion might render
this turbulent viscosity non-isotropic, with important effects
on the transport of angular momentum within the convection zone.
Of course, such effects were not included in our derivation of the
Taylor-Proudman theorem, where the inviscid fluid equations were used
and the flow was assumed to be described just by the (large-scale) 
rotation.  In contrast, simple models of convection-zone dynamics, with 
parametrized anisotropic viscosity, have in fact had some success
in reproducing the helioseismically inferred rotation rate
(Pidatella et al.\ 1986).

\begin{figure}
\begin{center}
\leavevmode\epsfxsize=9.5cm \epsfbox{\fig/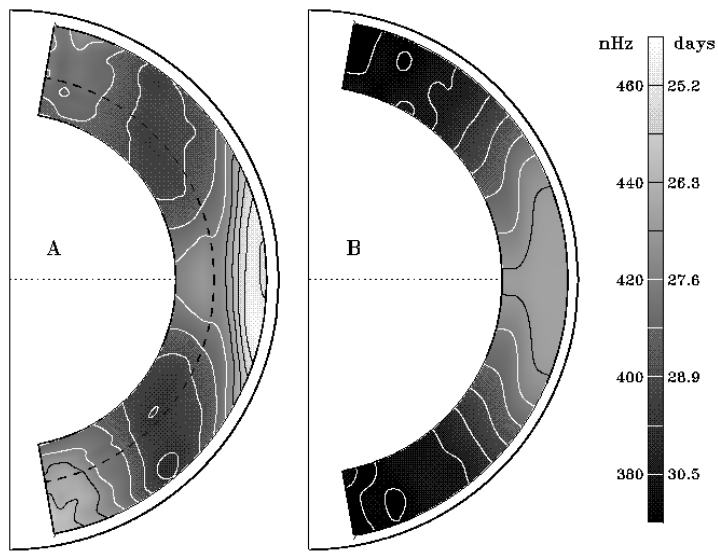} 
\end{center}
\caption{
\dblspace
The results of two simulations of convection-zone rotation by Miesch, Elliott
and Toomre (Miesch 2000). The two simulations
use different boundary conditions and parameter values, and 
illustrate some of the range of possible responses of the differential 
rotation to the form of the convection.  Also note that Simulation~A
has a higher resolution and includes penetration into a stable region
beneath the convection zone, whereas the convective motions in Simulation~B
are more laminar and there is no penetration beneath the convection zone.
}
\label{fig:fig6}
\end{figure}

Recent advances in computing power have led to improved numerical
simulations (Miesch et al.\ 2000; Brun \& Toomre\ 2001),
which come closer to representing turbulent 
convective flow regimes such as exist in the Sun's convection zone. 
Figure~6 shows results from two such simulations by Miesch, 
Elliott and Toomre. The simulations can yield a range of 
differential rotation profiles, depending on the conditions imposed
at the top and bottom boundaries of the simulation region, and on
the parameter values adopted for the problem. Since it is not obvious
what are the most appropriate boundary conditions and parameter
values to choose, it is necessary to explore various possibilities and
study the different responses. Simulation~B
has rotation contours at mid-latitudes which
are nearly radial, as in the Sun (compare Fig.~\ref{fig:fig3}), 
but the contrast in rotation rate
between low and high latitudes is not as great as is observed in
the Sun (about 70 nHz, rather than 130 nHz).
In case~A, the latitudinal variation of the Sun's rotation is better
reproduced, but the mid-latitude contours do not look quite as similar
to those in Fig.~\ref{fig:fig3}. Nonetheless these results are encouraging 
indications that
we may be close to reproducing theoretically the gross features
of the solar rotation inferred by helioseismology. There is still 
though much work ahead, both observational and theoretical, in 
getting a detailed understanding of the Sun's rotation and with that,
we hope, a better understanding of the solar activity cycle and
of large-scale rotating fluid systems on planets and stars.

\section{Final remarks}

The advent of helioseismology has revolutionized our knowledge of the 
Sun's internal rotation, showing that the dynamics of the solar interior
are quite different from the pre-helioseismic model simulations. The 
rotation within the convection zone is now well established except 
perhaps at high latitudes, and the more turbulent simulations appear
to be getting close to matching observation. The existence of the 
tachocline and the near-uniform rotation in much of the deeper
interior are also well established, and suggest that a 
magnetic field enforces essentially rigid rotation in at least the
outer part of the Sun's radiative interior. The rotation of the very deepest
part of the Sun is still somewhat uncertain but a very fast-spinning 
core is ruled out unless it is very small. The temporal variations of
the near-surface banded flows are firmly established but their 
origin is still incompletely understood. The less certain aspects of the 
helioseismic inferences are the temporal variations around the tachocline
region and the high-latitude jet: we must wait to see
whether these are confirmed by future observations. 

In addition to the global helioseismic investigations, the newer field 
of local helioseismology (described more fully in the chapter by Kosovichev)
is providing knowledge of the dynamics of the near-surface layers of the 
Sun. These techniques, which by virtue
of their local character can also provide information about how
flows and structure vary with longitude and in the northern and southern
hemispheres,
reveal not only the rotational flow field, but
also meridional (i.e., North-South) flows. Meridional flows are likely to 
play a significant role in understanding the angular-momentum budget
in the convective envelope. Recent findings with one such local
helioseismic technique (Haber et al. 2001) 
indicate fascinating changes in the structure
of the meridional circulation patterns in one hemisphere of the Sun
from year to year; also, they confirm the migrating banded flows in the 
near-surface rotation but indicate that they can be markedly
asymmetric about the equator. These near-surface findings, combined with
the deeper inferences of the global methods, continue to provide an 
intriguing insight into the dynamics of the solar interior.

\bigskip\noindent
{\it Acknowledgements} We are grateful 
to Prof. R. K. Ulrich for providing Fig.~\ref{fig:surf_rot}, 
to Dr. R. Howe providing Figs~\ref{fig:zonal} and \ref{fig:tachoosc},
and to Dr. M. Miesch for providing Fig.~\ref{fig:fig6}.
We are happy to acknowledge the financial support of the 
UK Particle Physics and Astronomy
Research Council, and the Danish National Research Foundation through its
establishment of the Theoretical Astrophysics Center.
Our thanks also go to Prof. Juri Toomre and Dr. Michael Kn\"olker for
hospitality at JILA and HAO respectively, during the time 
when much of this chapter was written.

\end{document}